\documentclass[aps,prb,twocolumn,groupedaddress,showpacs,floatfix]{revtex4}
\usepackage{graphicx}
\begin{document}

\title{Gradient-dependent density functionals of the PBE type for atoms,
molecules and solids}
\author{Luana S. Pedroza and Antonio J. R. da Silva}
\affiliation{Instituto de F\'{\i}sica, Universidade de S\~ao Paulo,
Caixa Postal 66318, S\~ao Paulo, 05315-970, SP, Brazil}
\author{K. Capelle}
\affiliation{Instituto de F\'{\i}sica de S\~ao Carlos,
Universidade de S\~ao Paulo, Caixa Postal 369, S\~ao Carlos,
13560-970 SP, Brazil}
\date{\today}

\begin{abstract}
One of the standard generalized-gradient approximations (GGAs) in
use in modern electronic-structure theory, PBE, and a recently
proposed modification designed specifically for solids, PBEsol,
are identified as particular members of a family of functionals
taking their parameters from different properties of homogeneous
or inhomogeneous electron liquids. Three further members of this
family are constructed and tested, together with the original PBE
and PBEsol, for atoms, molecules and solids. We find that PBE, in
spite of its popularity in solid-state physics and quantum
chemistry, is not always the best performing member of the family,
and that PBEsol, in spite of having been constructed specifically
for solids, is not the best for solids. The performance of GGAs
for finite systems is found to sensitively depend on the choice of
constraints steaming from infinite systems. Guidelines both for
users and for developers of density functionals emerge directly
from this work.
\end{abstract}



\maketitle

\newcommand{\be}{\begin{equation}}
\newcommand{\ee}{\end{equation}}
\newcommand{\bea}{\begin{eqnarray}}
\newcommand{\eea}{\end{eqnarray}}
\newcommand{\bi}{\bibitem}

Modern density-functional theory (DFT)\cite{kohnrmp,dftbook,parryang} owes its
success and popularity largely to the availability of simple and reliable
density functionals.\cite{perdewjcp} Among the most widely used such
functionals are gradient-dependent approximations, such as the B88 exchange
functional\cite{b88} or the PBE generalized-gradient approximation (GGA)
for exchange and correlation.\cite{pbe} Although many other functionals
are available, PBE is today the {\em de facto} standard for gradient-dependent
functionals in solid-state physics, and, together with B88, in quantum
chemistry. These gradient-dependent functionals also form the basis for the
development of more sophisticated functionals, of, {\em e.g.} the meta-GGA
or hybrid type.

Given the importance of PBE both for countless practical
applications of DFT, as well as for constructing more refined
approximations, it is not surprising that over the years many
variations of the basic PBE form have been
developed.\cite{revPBE,RPBE,mPBE,xPBE,wc} Most of these are more
empirical than the original construction, in the sense that they
include parameters fitted to test sets of selected systems and
properties. None is uniformly better than the original PBE for all
systems and properties. For more restricted classes of systems,
however, it is not that hard to improve on PBE, as is illustrated
by the recent proposal of PBEsol,\cite{pbesol} which was designed
to improve on PBE specifically for solids. As this is still an
extraordinary large and diverse class of systems, even an
improvement `only' for solids is still a very major step forward,
and consequently PBEsol is currently being implemented in many
standard electronic-structure codes, and intensely
scrutinized.\cite{pbesolrefs,haas}

In the present paper we point out that the step that led from PBE to PBEsol
is, in fact, not unique, and allows several variations. We propose a family
of functionals, which we call PBE($\beta$,$\mu$), of which both the original
PBE and the original PBEsol are particular members, and which includes at
least three more alternatives. Each member of this family takes the value
of the $\beta$ and $\mu$ parameter from a different physical constraint,
without statistical fitting to test sets of systems. Five PBE($\beta$,$\mu$)
functionals are systematically tested for atoms, molecules and solids, and,
within each class, for systems with physically and chemically different
properties.

\begin{table*}
\begin{ruledtabular}
\caption{\label{table1} Specification of the values of $\beta$ and
$\mu$ used in each of the investigated variants of PBE$(\beta,\mu)$.}
\begin{tabular}{ccccccc}

& PBE$(G_c,J_r)$  & PBE$(J_s,G_x)$ & PBE$(J_s,J_r)$ & PBE$(G_c,G_x)$ & PBE$(J_r,G_x)$  & PBE$(G_c,J_s)$ \\

& = PBE           & = PBEsol & new & new & new  & new\\

$\beta$ & 0.066725        & 0.046  & 0.046          & 0.066725 & $3\mu/\pi^2$
& 0.066725 \\


$\mu$   & $\pi^2 \beta/3$ & 10/81  & $\pi^2 \beta/3$ & 10/81    & 10/81 &
n.a. \\

\end{tabular}
\end{ruledtabular}
\end{table*}

To begin, we recall that the original PBE functional has three
parameters. One, $\beta$, appears in the correlation functional,
and was originally obtained by requiring that for weakly
inhomogeneous high-density systems the second-order gradient
expansion of the correlation energy was
recovered.\cite{mabrueckner} Another, $\mu$, appears in the
exchange functional, and in the original construction was obtained
from requiring that the combined exchange and correlation
functional predicts the correct linear-response of bulk
Jellium.\cite{pbe} This latter requirement implies $\mu = \pi^2
\beta /3$. In PBE, this is used to fix $\mu$, once $\beta$ is
obtained from the correlation energy gradient expansion. 
(One of the alternatives we propose here is to invert this relation, 
using it to determine $\beta$ once $\mu$ is obtained from the gradient
expansion of the exchange energy.)

In PBEsol, $\beta$ is determined instead by fitting to Jellium
surface energies (JSEs) (a strategy proposed earlier by Armiento
and Mattsson\cite{am05}), whereas $\mu$ is chosen such as to
recover the second-order gradient expansion of the exchange
functional. Our original motivation for reconsidering the choice
of $\beta$ and $\mu$ was that the JSEs used for obtaining $\beta$
in PBEsol had themselves been calculated from the TPSS meta-GGA
functional,\cite{tpss} whose correlation energy in turn contains
as an ingredient the original PBE correlation. It seems slightly
inconsistent to us to employ in the construction of a functional
designed to improve on PBE, a fit to data obtained from a
functional that itself contains PBE. In practice, however, this
may be a rather purist objection, as the TPSS JSEs are quite close
to those obtained by other methods.\cite{tpsssurfaceenergies,wood}

A search for possibilities to avoid this small inconsistency,
however, suggested a broader perspective on PBE and PBEsol: since
the aim of PBEsol was to improve on PBE specifically for solids,
and the gradient expansion is more relevant for the slowly varying
densities of typical solids than it is for molecular densities, it
seems promising to take {\em both} parameters, $\beta$ and $\mu$,
from the gradient expansion, without making any use of properties
of Jellium. An additional advantage of extracting both parameters
from the same source is that it enhances the chances of error
compensation between the exchange and the correlation functional
-- an effect that is known to be behind the success of the
local-density approximation, and that has only partially been
preserved by common GGAs.

Further exploration of the idea of a consistent (in the sense of
coming from the same source) set of parameters suggests the
alternative possibility to take both parameters, $\beta$ and
$\mu$, from properties of Jellium, without making any use of
gradient expansions. Specifically, we can determine $\beta$ from
JSEs and $\mu$ from the Jellium response function. {\em A priori},
this choice, too could be expected to be good for solids, or at
least for metals, as Jellium is the paradigm of metallic behavior
in extended systems.

Table~\ref{table1} presents a summary of the different variations
of PBE and the corresponding values of $\beta$ and $\mu$. Our
notation for the entire family of functionals is PBE$(\beta,\mu)$,
where $\beta$ and $\mu$ are replaced by $G_x$ ($G_c$) or $J_s$
($J_r$), depending on whether the parameter has been obtained from
the gradient expansion for exchange (correlation), or from Jellium
surface (response properties). In this notation the original PBE
becomes PBE$(G_c,J_r)$, the original PBEsol becomes
PBE$(J_s,G_x)$, and the two alternatives just described read
PBE$(G_c,G_x)$ and PBE$(J_s,J_r)$.

Table~\ref{table1} also includes a further mixed choice, in which
$\beta$ is determined from the Jellium response function and $\mu$
from the gradient expansion of the exchange energy, {\em i.e.}
PBE$(J_r,G_x)$. This is the functional alluded to above, in
connection with inversion of the relation $\mu=\pi^2 \beta/3$.
Mathematically, still another possibility would be to determine
$\beta$ from the gradient expansion of the correlation energy and
$\mu$, {\em e.g.}, from fitting to JSEs, resulting in
PBE$(G_c,J_s)$, but the results from that fit are not available,
so we cannot test that option here.

We note that PBE and PBEsol contain a third parameter, $\kappa$,
which is determined from the Lieb-Oxford (LO)
bound\cite{lieboxford}. Recently it has been suggested that the
value of this parameter can be readjusted to possible tighter forms of
the LO bound.\cite{refsLO,truhar08}
In the present work we do not change $\kappa$ relative to the
original PBE proposal, in order not to mix the question of a
possible tightening of the LO bound with that of consistent
choices of the $\beta$ and $\mu$ parameters. Future work should
explore the consequences of combined changes of all three
parameters. The proposal of Ref.~\onlinecite{truhar08}, which in
addition to values of parameters such as $\kappa$, $\beta$ and
$\mu$ also changes the form of the PBE enhancement factor, is a
step in this direction.

Restricting ourselves to exploring changes of only $\beta$ and $\mu$, we
report in Tables~\ref{table2} to \ref{table4} the mean error, defined 
as $me = \frac{1}{N} \sum_{i=1}^{N}{(R_i-A_i)} $; 
the mean absolute error ($mae = \frac{1}{N} \sum_{i=1}^{N}
\left|{R_i-A_i}\right|$)
and the mean absolute relative error ($mare = \frac{1}{N} \sum_{i=1}^{N}
\left|(R_i-A_i)/R_i\right|$),
where $R_i$ is the reference (benchmark) value of the $i$'th system and
$A_i$ the corresponding approximate value, for $N=26$ light
atoms,\cite{footnote0} $N=14$
molecules\cite{footnote1} and $N=13$ extended systems,\cite{footnote2}
chosen according to two requirements: (i) reliable
benchmark data are available, and (ii) different types of crystal structures
and of chemical bonds are represented in the data set. 
In the following analysis, we focus mainly on the trends of the mare. 

\begin{table}
\begin{ruledtabular}
\caption{\label{table2} Mean error (me), mean absolute error (mae)
and mean absolute relative error (mare) (same labels are used in
all Tables) for ground-state energies (GSE) $E_0$ (Ry) and ionization
potentials $IP$ (Ry) for 26 light atoms\cite{footnote0} ($Z=3-28$); and
relative percentage error for the H-atom GSE (RE$_H$).}
\begin{tabular}{ccccccc}

  &PBE$(\beta,\mu)$ & $G_c,J_r$ & $J_s,G_x$ & $J_s,J_r$ & $G_c,G_x$ & $J_r,G_x$ \\
  &me           & 9.052     & 7.790     & 8.231     & 7.594     & 7.893     \\
$E_0$  &mae      & 9.354     & 9.210     & 9.232     & 9.207     & 9.211     \\
  &mare             & 0.00451   & 0.00627   & 0.00560   & 0.00662   & 0.00609    \\
RE$_H$  &  \%          & 0.00191   & 2.26      & 1.43      & 2.65      & 2.03       \\
\hline
   &me               & -0.245    & -0.247    & -0.244    & -0.254    & -0.243    \\
IP &mae              &  0.258    &  0.260    &  0.257    &  0.266    &  0.257    \\
   &mare             &  0.368    &  0.370    &  0.367    &  0.380    &  0.366

\end{tabular}
\end{ruledtabular}
\end{table}

\begin{table}
\begin{ruledtabular}
\caption{\label{table3} The me, mae and mare for atomization
energies AE (eV) and interatomic distances $d$ (\AA) for 14 small
molecules. 
\cite{footnote1}}
\begin{tabular}{ccccccc}
   &PBE$(\beta,\mu)$ & $G_c,J_r$ & $J_s,G_x$ & $J_s,J_r$ & $G_c,G_x$ & $J_r,G_x$ \\
   &me               &  0.0452   & 0.566     & 0.532     & 0.458     & 0.615     \\
AE &mae              &  0.299    & 0.614     & 0.584     & 0.616     & 0.647     \\
   &mare             &  0.0573   & 0.100     & 0.105     & 0.112     & 0.103     \\

\hline
  &me               & -0.0143   & -0.0122    & -0.0115  & -0.0124   & -0.00959  \\
$d$ &mae              &  0.0231   &  0.0214    &  0.0201  &  0.0307   &  0.0223   \\
  &mare             &  0.0156   &  0.0152    &  0.0146  &  0.0204   &  0.0156
\end{tabular}
\end{ruledtabular}
\end{table}

For atoms, we have performed all-electron calculations using the
mesh-based (basis-set free) atomic code that is part of the Siesta
package.\cite{siesta} We find that the original PBE predicts best
ground-state energies. The alternative functional PBE$(J_s,J_r)$,
which takes both of its parameters from Jellium properties, comes
second. This is rather unexpected, as JSEs (used to obtain
$\beta$) and bulk Jellium response functions (used to obtain
$\mu$) are maximally different from what one expects for atomic
densities. Probably, this indicates a substantial error
cancellation between exchange and correlation, made possible by
taking both parameters from the same reference system.

\begin{table}
\begin{ruledtabular}
\caption{\label{table4} The me, mae, and mare for lattice
constants $a$ (\AA), bulk moduli $B$ (GPa) and cohesive energies CE (eV) for 13
solids.\cite{footnote3}}
\begin{tabular}{ccccccc}
  & PBE$(\beta,\mu)$ & $G_c,J_r$ & $J_s,G_x$ & $J_s,J_r$ & $G_c,G_x$ & $J_r,G_x$ \\
  & me               & -0.129   & -0.0722    & -0.0881   & -0.0692   & -0.0724   \\
$a$ & mae              &  0.129   &  0.0739    &  0.0890   &  0.0692   &  0.0753   \\
  & mare             &  0.0283  &  0.0165    &  0.0198   &  0.0153   &  0.0168   \\

\hline
   & me              & 35.81     & 25.65     & 29.57     & 24.06     &  26.80    \\
$B$ & mae             & 35.99     & 25.84     & 29.76     & 24.21     &  27.03    \\
   & mare            & 0.244     & 0.173     & 0.201     & 0.153     &  0.184     \\

\hline
   & me               & -0.521    &  0.00831  & -0.172    &  0.0136   & -0.0136   \\
CE & mae              &  0.555    &  0.282    &  0.280    &  0.288    &  0.278    \\
   & mare             &  0.117    &  0.0604   &  0.0659   &  0.0644   &  0.0598

\end{tabular}
\end{ruledtabular}
\end{table}

For atoms we also compared Koopman's theorem ionization
potentials, {\em i.e.} the negative of the eigenvalue of the
highest occupied Kohn-Sham orbital. Here original PBE,
PBE$(J_s,J_r)$ and the novel mixed choice PBE$(J_r,G_x)$ perform
very similarly, whereas PBEsol and PBE$(G_c,G_x)$ are clearly
worse. However, as ionization energies obtained from eigenvalues
suffer from the self-interaction error and the resulting wrong
asymptotics of the effective potentials, which is not addressed at
all by changing the values of $\beta$ and $\mu$, a comparison of
the performance of PBE$(\beta,\mu)$ for these quantities is less
conclusive than that for total energies.

As a final test for atoms, we compared the error in the ground-state
energy of the hydrogen atom. This error, arising in an $N=1$ electron 
system, is exclusively due to self-interaction. As the fifth line of
Table~\ref{table2} shows, the original PBE functional performs by far best
of all tested variants, distantly followed by PBE$(J_x,J_r)$. As all
constraints investigated here stem from infinite systems, this illustrates
a dramatic sensitivity of the performance of approximate functionals near
$N=1$ to changing constraints arising from $N\to\infty$.

For molecules and solids, our calculations were performed with the
SIESTA code,\cite{siesta} employing norm-conserving
Troullier-Martins pseudopotentials and a strictly localized TZP and
DZP basis set for molecules and solids, respectively. Pseudopotentials
for use with any of the PBE$(\beta,\mu)$ functionals were generated by
employing the same functional also in the atomic all-electron
calculations. Convergence with respect to the basis set size was
checked by repeating the molecular PBE calculations, as well as
selected atomic and solid-state calculations, independently
with a plane-wave code that is part of the CPMD
package,\cite{cpmd} where convergence is controlled by a single
parameter, the energy cutoff.

For molecules, original PBE=PBE$(G_c,J_r)$ predicts best
atomization energies, whereas the alternative PBE$(J_s,J_r)$
predicts best interatomic distances. For both quantities
PBEsol=PBE$(J_s,G_x)$ comes second.

For solids, the original PBE functional is the worst of all tested
members of the PBE$(\beta,\mu)$ family. Not surprisingly, PBEsol
consistently outperforms PBE, but interestingly, it is itself
outperformed by PBE$(G_c,G_x)$ for lattice constants and bulk
moduli, and by the novel mixed choice PBE$(J_r,G_x)$ for cohesive
energies. Overall, the best performance is obtained from
PBE$(G_c,G_x)$.\cite{footnote3,footnote3b}

The present results suggest guidelines for the informed choice of
density functionals and their construction: (i) The original PBE
is not optimal for use in solid-state calculations and for
molecular bond lengths. (ii) PBEsol, in spite of its name, is not
the best PBE-type functional for bulk properties of solids, where
PBE$(G_c,G_x)$ performs much better. Surfaces require separate
analysis. (iii) A consistent choice of parameters (in the sense of
stemming from the same type of source) tends to benefit the
performance of the functional: PBE$(G_c,G_x)$ and PBE$(J_s,J_r)$
are more often among the best or second-best performers than any
of the other variants.\cite{footnote4} (iv) The good performance
of PBE$(G_c,G_x)$ for solids suggests that restoring the gradient
expansion completely (for exchange and correlation, and not just
partially, for exchange only, as in PBEsol) is beneficial for
extended systems, whereas for finite systems (atoms and molecules)
gradient expansions turns out to be a less important ingredient.
(v) A common ingredient of functionals that perform well for
finite systems is the Jellium response function, but there seems
to be no systematic trend for the utility of the Jellium surface
energy. (vi) Substituting a constraint arising from infinite
systems by another, also arising from infinite systems, can have a
dramatic effect even at the opposite end of the spectrum, for $N=1$.

Users of PBE and related functionals should thus be aware of the fact that
within the functional form of PBE there is not one single choice of
constraints that benefits all system types. A related aspect is that the
original PBE has been used as ingredient in the
development of recent meta-GGA functionals, such as TPSS\cite{tpss} and
PKZB.\cite{pkzb} These constructions should also be reconsidered in view of
the superior performance of some of the other PBE$(\beta,\mu)$ variants, as
more sophisticated high-level functionals should naturally be build using
as ingredients the most reliable functionals from lower levels.

{\it Acknowledgments}
This work was supported by FAPESP and CNPq.
We thank M. M. Odashima for useful remarks on an earlier version of this 
manuscript.

\end{document}